\documentclass{aa}

\usepackage{graphicx}
\usepackage{natbib}
\usepackage{txfonts}
 
\begin{document}

\title {Effects of the radial flows on the chemical evolution of the
  Milky Way disk }

\author {E. Spitoni\inst{1,3} \thanks {email to:
    spitoni@oats.inaf.it}\and F. Matteucci\inst{1,2}} \institute{
  Dipartimento di Astronomia, Universit\`a di Trieste, via
  G.B. Tiepolo 11, I-34131, Trieste, Italy \and I.N.A.F. Osservatorio
  Astronomico di Trieste, via G.B. Tiepolo 11, I-34131, Trieste,
  Italy \and Department of Mathematics, University of \'Evora, R. Rom\~ao
Ramalho 59, 7000 \'Evora, Portugal}

\date{Received xxxx / Accepted xxxx}

\abstract{The majority of chemical evolution models assume that the
  Galactic disk forms by means of infall of gas and divide the disk
  into several independent rings without exchange of matter between
  them. However, if gas infall is important, radial gas flows should
  be taken into account as a dynamical consequence of infall.} {The
  aim of this paper is to test the effect of radial gas flows on
  detailed chemical evolution models (one-infall and two-infall) for
  the Milky Way disk with different prescriptions for the infall law
  and star formation rate.}  {We modified the equation of chemical
  evolution to include radial gas flows according to the method
  described in Portinari \& Chiosi} {We found, that with a gas radial
  inflow of constant speed the metallicity gradient tends to
  steepen. Taking into account a constant time scale for the infall
  rate along the Galaxy disk and radial flows with a constant speed,
  we obtained a too flat gradient, at variance with data, implying
  that an inside-out formation and/or a variable gas flow speed are
  required.  To reproduce the observed gradients the gas flow should
  increase in modulus with the galactocentric distance, both in the
  one-infall and two-infall models. However, the inside-out disk
  formation coupled with a threshold in the gas density (only in the
  two-infall model) for star formation and/or a variable efficiency of
  star formation with galactocentric distance can also reproduce the
  observed gradients without radial flows.}{We showed that the radial
  flows can be the most important process in reproducing abundance
  gradients but only with a variable gas speed. Finally, one should
  consider that uncertainties in the data concerning gradients prevent
  us to draw firm conclusions. Future more detailed data will help to
  ascertain whether the radial flows are a necessary ingredient in the
  formation and evolution of the Galactic disk and disks in general.}


\titlerunning{Effects of the radial flows   }
\authorrunning{Spitoni et al.}
\maketitle
 
\section{Introduction}

Generally, a good agreement between model predictions and observed
properties of the Galaxy is obtained by models which assume that the
disk formed by infall of gas: (e.g Matteucci \& Fran\c cois (1989),
Chiappini et al. 1997, Fran\c cois et al. 2004).



The formation timescale of the thin disk is assumed to be a
function of the galactocentric distance, leading to an inside-out
scenario for the Galaxy disk build-up (Matteucci \& Fran\c cois
1989). This assumption helps in reproducing the abundance gradients
along the disk as shown in studies such as  Chiappini et al. (2001), Cescutti
et al. (2007), Colavitti et al. (2009) and Marcon-Uchida et
al. (2010). In these papers were analyzed the parameter space of the
``static'' chemical evolution models considering different infall and
star formation (SF) rate laws in order to reproduce the observed
abundance gradients.  Cescutti et al. (2007) showed that the results
obtained using a chemical evolution model with a two-infall law and
inside-out formation are in very good agreement with the data on
Cepheids in the galactocentric distance range 5-17 kpc for many of the
studied elements.

Colavitti et al. (2009) found that it is impossible to fit all the
disk constraints at the same time without assuming an inside-out
formation for the Galactic disk together with a threshold in the gas
density for the SF rate. In particular, the inside-out formation is
important to reproduce the right slope of the abundance gradients in
the inner disk, whereas the threshold gas density is important to
reproduce present day gradients in the outer disk. Generally, simple
models without radial gas flows and with a radially constant formation
timescale cannot reproduce the slope in the inner disk. In the
framework of models with no inside-out mechanism such as the model
with the cosmological infall law of Colavitti et al. (2009) they
tested the effect of the efficiency of SF varying with the
galactocentric radius, being higher in the innermost than in the
outermost regions of the Galactic disk. This assumption, even without
the threshold in the surface gas density, can produce gradients with
the right slope both in the inner and outer regions of the Galactic
disk but it fails in reproducing the gas density distribution along
the disk.  However, if gas infall is important, the Galactic disk is
not adequately described by a simple multi-zone model with separated
zones (Mayor \& Vigroux 1981). To maintain consistency, radial gas
flows have to be taken into account as a dynamical consequence of
infall.  The infalling gas has a lower angular momentum than the
circular motions in the disc, and mixing with the gas in the disc
induces a net radial inflow.  Lacey \& Fall (1985) estimated that the
gas inflow velocity is up to a few km s$^{-1}$.  Goetz \& K\"oppen
  (1992) studied numerical and analytical models including radial
  flows. They concluded that radial flows alone cannot explain the
  abundance gradients but are an efficient process to amplify the
  existing ones. In particular, they suggested that the speed of the
  flow should increase with galactocentric distance and only change by
  0.15 km s$^{-1}$ per kpc, well below any detection limit for drift
  velocities of molecular clouds.   Portinari \& Chiosi (2000)
implemented in the detailed chemical evolution model of Portinari \&
Chiosi (1999), characterized by a single infall episode, the effects
of radial gas inflows by means of the following assumptions:
\begin{itemize}
\item

an uniform inflow pattern  through the disk, with speed
v$_R=-1$~km s$^{-1}$ applied to a model with a Schmidt (1959) SF law and they 
found , as expected, that the
metallicity gradient tends to steepen in the presence of radial gas
flows;
\item 
they considered models with various SF laws and tuned the inflow
velocity pattern, in each case,  as to match the observational data
relative to the radial profiles of the Galactic disk.  They suggested
that inclusion of radial flows in the chemical models can improve the
match with the data. For these models the radial velocities span the
range $-1 \leq$ v$_R$ $\leq -0.3$ km s$^{-1}$.

\end{itemize}
 In the work of Sch\"onrich \& Binney (2009) 
   both the stellar and gas flow were considered in a chemical evolution model without
   inside-out formation. They are able to reproduce the observed data
   considering a variable speed of the gas flow in the range 0 km
   s$^{-1}$ to -5 km s$^{-1}$.

Another physical cause of radial flows is the gas viscosity. Viscosity
in the gas layers induces radial inflows in the inner parts of the
disc and outflows in the outer parts, with velocities of $\sim$0.1~
km~ s$^{-1}$ (Lacey \& Fall 1985). Thon \& Meusinger (1998) presented
a model for the chemical and dynamical evolution of the Galactic disk
which combines viscous radial gas flows with infall of external gas
onto the disk, and infall-induced radial gas flows. In this
hydrodynamic ``hybrid'' model it was assumed that the viscosity of the
gas is closely related to the processes of SF and the
energetic response from massive stars, i.e. the timescales for viscous
transport and for gas consumption by SF are nearly the
same. Following the parameterization of Sommer-Larsen \& Yoshii
(1989), they obtained a SF rate law depending on the viscosity parameters.

Gravitational interactions between gas and spiral density waves lead
to large-scale shocks, dissipation and therefore radial inflows of gas
(or outflows in the outer parts) with typical velocities of
$\sim$0.3~km~s$^{-1}$ (e.g. Bertin \& Lin 1996 and references
therein); much larger velocities can be achieved in the inner few kpc
in the presence of a barred potential. However, given the small values
of the flow speed in these works, the effect on the abundance
gradients are negligible.

In this work we include radial inflows in 2 different chemical
evolution models: a one-infall and a two-infall model.  In particular, this is the first time that radial flows are included in the two-infall model.
We test a
constant velocity pattern for the radial inflow as well as a best fit
pattern as done in  Portinari \& Chiosi (2000).  We then study the effect
of radial flows on the formation of gradients coupled with the
inside-out mechanism formation and the effect of the threshold in the
SF.  Last but not least, we compare our results with the most recent data
now available, including Cepheids,and this represents a novelty relative to previous works. This comparison will allow us to put constraints on the mechanisms of gradient formation.

The paper is organized as follow: in Sect. 2 we describe the reference
``static'' models used in this work, in Sect. 3 we report the
nucleosynthesis prescriptions, in Sect. 4 we present the
implementation of the radial inflow of gas in a detailed chemical
evolution model, in Sect. 5 observational data are shown. In Sect. 6
we report and discuss our results. Finally we draw the main
conclusions in Sect. 7.

\section{The chemical evolution model for the Milky Way}

In order to reproduce the chemical evolution of the thin-disk, we
adopt as a reference model an updated one-infall version of the chemical
evolution model presented by Matteucci \& Fran\c cois (1989). In this model,
the galactic disk is divided into several concentric rings which
evolve independently without exchange of matter.

The infall law for the thin-disk is defined as: 
 
\begin{equation}\label{infall} 
B(r,t)= b(r)e^{-\frac{t}{\tau_D}}, 
\end{equation} 
 
where $\tau_D$ is the timescale for the infalling gas into the thin-disk.
It is worth noting that eq.(\ref{infall}) differs from the two-infall
model of Chiappini et al. (1997) which assumes an infall law such as:
 \begin{equation}
A(r,t)= a(r) e^{-t/ \tau_{H}(r)}+ b(r) e^{-(t-t_{max})/ \tau_{D}(r)}.
\end{equation}

To have an inside-out formation on the disk, the timescale for the mass
accretion is assumed to increase with the Galactic radius following a
linear relation given by (see Chiappini et al. 1997):
\begin{equation}
\tau_{D}=1.033 r (\mbox{kpc}) - 1.27 \,\, \mbox{Gyr},
\label{t1}
\end{equation}
  for galactocentric distances $\geq$ 4 kpc. The region within 4 kpc is not considered in this paper and has by the strict inflow no impact on the outer regions.  
The coefficient $b(r)$ is obtained  by imposing a fit  
to the observed current total surface mass density in the thin disk
as a function of the galactocentric 
distance given by:
\begin{equation}
\sigma(r)=\sigma_{D}e^{-r/r_{D}},
\end{equation}
where $\sigma_{D}$=531 $M_{\odot}$ pc$^{-2}$  is the central total surface mass density and
$r_{D}= 3.5$ kpc is the scale length.  
 
In order to make the program as simple and generalized as possible, we
used a SF rate proportional to a Schmidt (1959) law:
 
\begin{equation} 
\psi(r,t) \propto \nu \Sigma_{gas}^k(r,t) 
\end{equation}

where $\nu$ is the efficiency in the SF process and the
surface gas density is represented by $\Sigma_{gas}(r,t)$ while the
exponent $k$ is equal to 1.4 (see Kennicutt 1998).

The equation below describes the time evolution of $G_{i}$, namely the
mass fraction of the element $i$ in the gas:

\begin{displaymath}
\dot{G_{i}}(r,t)=-\psi(r,t)X_{i}(r,t)
\end{displaymath}
\smallskip
\begin{displaymath}
+\int\limits^{M_{Bm}}_{M_{L}}\psi(r,t-\tau_{m})Q_{mi}(t-\tau_{m})\phi(m)dm
\end{displaymath}
\begin{displaymath}
+A_{Ia}\int\limits^{M_{BM}}_{M_{Bm}}\phi(M_{B})\cdot\left[\int\limits_{\mu_{m}}^{0.5}f(\mu)\psi(r,t-\tau_{m2})Q^{SNIa}_{mi}(t-\tau_{m2})d\mu\right]dM_{B}
\end{displaymath}
\begin{displaymath}
+(1-A_{Ia})\int\limits^{M_{BM}}_{M_{Bm}}\psi(r,t-\tau_{m})Q_{mi}(t-\tau_{m})\phi(m)dm
\end{displaymath}
\begin{displaymath}
+\int\limits^{M_{U}}_{M_{BM}}\psi(r,t-\tau_{m})Q_{mi}(t-\tau_{m})\phi(m)dm
\end{displaymath}
\smallskip
\begin{equation}
+X_{A_{i}}B(r,t),
\label{evol}
\end{equation}
 where $G_{i}(r,t)=[\sigma_g(r,t) X_i(r,t)]/\sigma_A(r)$,
  $\sigma_g(r,t)$ is the surface gas density, and $\sigma_A(r)$ is the
  present-time total surface mass density. 
$X_{A_{i}}$ are the abundances in the infalling material, which is generally
assumed to be primordial.
 $X_{i}(r,t)$ is the
abundance by mass of the element $i$ and $Q_{mi}$ indicates the
fraction of mass restored by a star of mass $m$ in form of the element
$i$, the so-called ``production matrix'' as originally defined by
Talbot \& Arnett (1973). We indicate with $M_{L}$ the lightest mass
which contributes to the chemical enrichment and it is set at
$0.8M_{\odot}$; the upper mass limit, $M_{U}$, is set at
$100M_{\odot}$.

For the IMF, we use that of Scalo (1986), constant in time and space.
$\tau_{m}$ is the evolutionary lifetime of stars as a function of
their mass {\it m} (Maeder \& Meynet 1989).

The Type Ia SN rate has been computed following Greggio \& Renzini (1983)
and  Matteucci \& Greggio (1986) and it is expressed as:
\begin{equation}
R_{SNeIa}=A_{Ia}\int\limits^{M_{BM}}_{M_{Bm}}\phi(M_{B})\left[
  \int\limits^{0.5}_{\mu_{m}}f(\mu)\psi(t-\tau_{M_{2}})d\mu \right]
dM_{B}
\end{equation}
where $M_{2}$ is the mass of the secondary, $M_{B}$ is the total mass
of the binary system, $\mu=M_{2}/M_{B}$,
$\mu_{m}=max\left[M_{2}(t)/M_{B},(M_{B}-0.5M_{BM})/M_{B}\right]$,
$M_{Bm}= 3 M_{\odot}$, $M_{BM}= 16 M_{\odot}$. The IMF is represented
by $\phi(M_{B})$ and refers to the total mass of the binary system for
the computation of the Type Ia SN rate, $f(\mu)$ is the distribution
function for the mass fraction of the secondary:
\begin{equation}
f(\mu)=2^{1+\gamma}(1+\gamma)\mu^{\gamma}  
\end{equation}
with $\gamma=2$; $A_{Ia}$ is the fraction of systems in the
appropriate mass range, which can give rise to Type Ia SN events. This
quantity is fixed to 0.05 by reproducing the observed Type Ia SN rate
at the present time (Mannucci et al. 2005). Note that in the case of
the Type Ia SNe the ``production matrix'' is indicated with
$Q^{SNIa}_{mi}$ because of its different nucleosynthesis contribution
(for details see Matteucci \& Greggio 1986 and Matteucci 2001).

We also apply the effect of radial inflows of gas on the model 
described in Fran\c cois et al. (2004). 

We do not enter into details of
this model, but we just remind here that in this case the Galaxy is
assumed to have formed by means of two main infall episodes: the first
formed the halo and the thick disk, the second the thin
disk. (i.e. two-infall model) The typical time-scale for the formation of the
halo is 0.8 Gyr and the entire formation period for the halo-thick disk does not
last more than 2 Gyr. The time-scale for the thin disk is much longer,
7 Gyr in the solar vicinity, implying that the infalling gas forming
the thin disk comes mainly from the intergalactic medium and not only
from the halo (Chiappini et al. 1997).  Moreover, the formation
timescale of the thin disk is assumed to be a function of the
galactocentric distance, leading to an inside-out scenario for the
Galaxy disk build-up (eq. \ref{t1}).  The galactic thin disk is
approximated by several independent rings, 2 kpc wide, without
exchange of matter between them.

 The main characteristic of the two-infall model is an almost
 independent evolution between the halo and the thin disk (see also
 Pagel \& Tautvaisienne 1995).  A threshold gas density of
 $7M_{\odot}pc^{-2}$ in the SF process (Kennicutt 1989,
 1998, Martin \& Kennicutt 2001) is also adopted for the disk. We
 consider for the halo a constant surface mass  density as a function of the
 galactocentric distance at the present time equal to 17
 $M_{\odot}$ $pc^{-2}$ and a threshold  for the halo phase of 4
 $M_{\odot}$ $pc^{-2}$ as it was assumed for the Model B in Chiappini et
 al. (2001).

 The time-steps are small for the first 2 Gyr, of the order of
  $10^{5}$ - $10^{6}$ years, then they become larger according to the
  smaller variation of the variables and reach values of the order of 1 Gyr.

\section{Nucleosynthesis prescriptions}

For the nucleosynthesis prescriptions of the O and Fe we adopt those
suggested in Fran\c cois et al. (2004).  They compared theoretical
predictions for the [el/Fe] vs. [Fe/H] trends in the solar
neighborhood for the above mentioned elements and they selected the
best sets of yields required to best fit the data. 
In particular they found that the Woosley \& Weaver (1995) metallicity dependent yields of SNe II provide the best fit to the data.
 No modifications are required for
the yields of Ca, Fe, Zn and Ni as computed for solar chemical
composition. For oxygen, the best results are given by the Woosley \&
Weaver (1995) yields computed as functions of the metallicity.  For
the other elements, variations in the predicted yields are required to
best fit the data (see Fran\c cois et al. 2004 for details).

 Concerning O, the best agreement with the [O/Fe] vs. [Fe/H] and the
 solar O abundance, as measured by Asplund et al. (2009), is obtained
 by adopting the original Woosley \& Weaver (1995) yields from massive
 stars as functions of metallicity. The same is not true for Fe: it
 was already known from the paper of Timmes et al. (1995) that the Fe
 yields as functions of metallicity overestimate the solar Fe and many
 people use those yields divided by a factor of 2. Alternatively, one
 can use the yields for solar chemical composition for the whole
 galactic lifetime and the result is the same, and this is what we do here.

Concerning the yields from Type SNeIa, revisions in the theoretical
yields by Iwamoto et al.(1999) have been taken into account.The
prescriptions for single low-intermediate mass stars are by van den
Hoek \& Groenewegen (1997), for the case of the mass loss parameter,
which varies with metallicity (see Chiappini et al. 2003, model5).

\section{The implementation of the radial inflow}

Lacey \& Fall (1985) showed that several mechanisms could drive a radial
gas flow in the Galactic disk, We refer here to the possibility that
the specific angular momentum of the infalling gas differs from the
value for the circular motions in the disk. Therefore, mixing will
induce a radial flow. Lacey \& Fall (1985) suggested that a good
estimate for velocity of that gas at 10 kpc is v$_R= -$ 1 km s$^{-1}$.
We implement radial inflows of gas on our reference model following the
prescriptions described in  Portinari \& Chiosi (2000).

Let the $k$-th shell be defined by the galactocentric radius $r_k$,
its inner and outer edge being labeled as $r_{k-\frac{1}{2}}$ and
$r_{k+\frac{1}{2}}$.  Through these edges, gas inflow with velocity
v$_{k-\frac{1}{2}}$ and v$_{k+\frac{1}{2}}$, respectively. The flow
velocities are taken positive outward and  negative inward.

 Radial inflows, with a flux $F(r)$, contribute to alter 
the gas surface density in the $k$-th shell in the following way:
\begin{equation}
\label{dsigmarf1}
\left[ \frac{d \sigma_{g k}}{d t} \right]_{rf} = 
   - \frac{1}{\pi \left( r^2_{k+\frac{1}{2}} - r^2_{k-\frac{1}{2}} \right) }
   \left[ F(r_{k+\frac{1}{2}}) - F(r_{k-\frac{1}{2}}) \right]
\end{equation}
where the gas flow at $r_{k+\frac{1}{2}}$ can be written as:
\begin{small}
\begin{equation}
\label{flux3}
F(r_{k+\frac{1}{2}}) = 2 \pi r_{k+\frac{1}{2}} \, v_{k+\frac{1}{2}} \left[\sigma_{g (k+1)} \right].
\end{equation}
\end{small}

As in  Portinari \& Chiosi (2000) we take the inner edge of the
$k$-shell, $r_{k-\frac{1}{2}}$, at the midpoint between the
characteristic radii of the shells $k$ and $k-1$, and similarly for
the outer edge $r_{k+\frac{1}{2}}$:

$r_{k-\frac{1}{2}}= (r_{k-1}+ r_k)/2$, and $r_{k+\frac{1}{2}}= (r_{k}+
r_{k+1})/2$. We get:
\begin{equation}
\left(r^{2}_{k+\frac{1}{2}} -r^{2}_{k-\frac{1}{2}} \right)=\frac{ r_{k+1}-r_{k-1}}{2}\left( r_k+\frac{r_{k-1}+r_{k+1}}{2}\right).  
\label{rs}
\end{equation}
Inserting
eqs. (\ref{flux3}) and (\ref{rs}) into  eq. (\ref{dsigmarf1}), we obtain the radial flow term to be added in eq. (\ref{evol}):

\begin{equation}
\left[ \frac{d}{dt} G_i(r_k,t) \right]_{rf} = -\, \beta_k \,
G_i(r_k,t) + \gamma_k \, G_i(r_{k+1},t),
\end{equation}

where $\beta_k$ and $\gamma_k  $ are, respectively:

\begin{equation}
\beta_k =  - \, \frac{2}{r_k + \frac {r_{k-1} + r_{k+1}}{2}}  
	\times \left[ v_{k-\frac{1}{2}}
	    \frac{r_{k-1}+r_k}{r_{k+1}-r_{k-1}}  \right]  
\end{equation}

\begin{equation}
\gamma_k =  - \frac{2}{r_k + \frac {r_{k-1} + r_{k+1}}{2}} 
	     \left[ v_{k+\frac{1}{2}}
	     \frac{r_k+r_{k+1}}{r_{k+1}-r_{k-1}} \right] 
	     \frac{\sigma_{A (k+1)}}{\sigma_{A k}}.
\end{equation}

$\sigma_{A (k+1)}$ and $\sigma_{A k}$ are the actual density profile
at the radius $r_{k+1}$ and $r_{k}$, respectively.  This formulation
for the radial inflow follows the one used in  Portinari \& Chiosi (2000)
but we want to stress that in the quantity $\beta_k$, concerning
the term in square brackets, contains  $r_{k+1}-r_{k-1}$
instead of $r_{k+1}+r_{k-1}$ as reported in the work of Portinari \& Chiosi (2000), but certainly they used the right
expression in their calculations. We assume that there are  no flows
from the outer parts of the disk where there is no SF. In our
implementation of the radial inflow of gas, only the gas which resides
inside the Galactic disk within the radius of 18 kpc can move inward
by radial inflow.  Our model is not subject to  numerical instabilities up to 18 kpc. Therefore since we choose not to consider flows from galactocentric distances larger than 18 kpc, our results are not dependent on these instabilities.

\section{Observational data}

\begin{figure}
\includegraphics[width=0.45\textwidth]{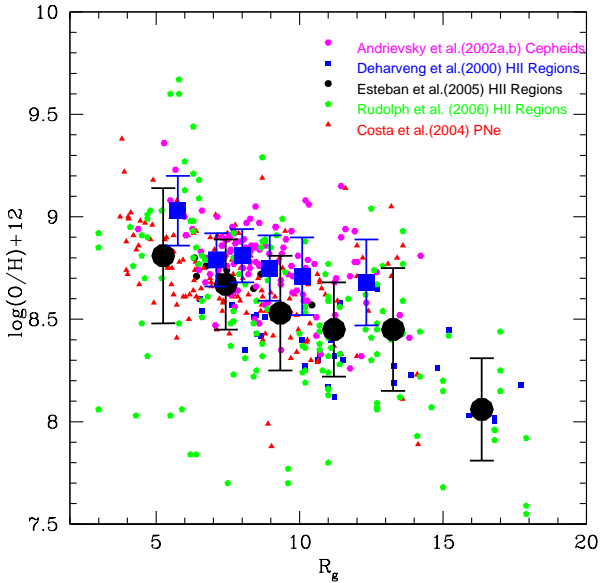}
\caption{Radial abundance gradient for oxygen from observations.  The data are taken by Deharveng et al. (2000)
  (filled blue squares), Rudolph et al. (2006) (filled green
  pentagon), Costa et al. (2004) (red filled triangles), Andrievsky et
  al. (2002 a,b) (magenta filled hexagons), and Esteban et al. (2005)
  (black filled circles).   We label with the large blue squared points the mean
  Cepheids values and relatives error bars, whereas with the large
  black filled circles the mean values form the HII and Planetary
  Nebulae regions. }
\label{dati}
\end{figure}

For the galactic abundance gradient of oxygen we use the following set
of data: Deharveng et al.  (2000), Esteban et al. (2005), Rudolph et
al. (2006), who analyzed the Galactic HII Regions; Costa et al. (2004)
who studied Planetary Nebulae (PNe); Andrievsky et al (2002a,b) who
analyzed Galactic Cepheids.  In order to compare
  our results with homogeneous sets of data, we separate the data
  by HII regions and PNe from the Cepheid ones.  To better understand
  the trend of the data, we divide the data in 6 bins as functions of
  the galactocentric distance. In each bin we compute the mean value
  and the standard deviations for the studied elements. In
  Fig. \ref{dati} we report the whole collection of the data used in
  this paper. We are aware that there  are some severe systematic uncertainties in each study that might justify larger error bars.

Concerning the data for the surface gas
density profile and the SF rate along the Galactic disk we use the
Dame et al. (1992) and Rana et al. (1991) ones.

\section{The model results}

\begin{figure}
\includegraphics[width=0.45\textwidth]{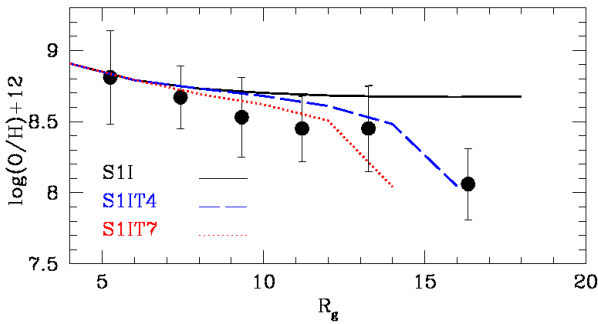}
\caption{Radial abundance gradient for oxygen. The black solid
  line refers to the one-infall model without threshold (S1I), the blue long
  dashed line to a one-infall model with a threshold of 4 $M_{\odot}
  pc^{-2}$ (S1IT4) and the red dotted line a model with a threshold of 7
  $M_{\odot} pc^{-2}$ (S1T7). The filled circles and relatives error bars are the observed values from HII regions and Planetary Nabulae. }
\label{tut}
\end{figure}

In this section the results of both the two-infall model of Chiappini
et al. (1997) and the one-infall model developed in this paper are presented. The
two infall model can have an effect on the disk evolution but only at
large galactocentric distances, as discussed in Chiappini et
al. (2001).

  We consider that in all models the efficiency of the SF
  is fixed at the value of $\nu=1$ Gyr$^{-1}$ in agreement with
  previous papers and in order to minimize the variation of model
  parameters.

 In this paper we will focus on the effect of radial flows
  with different prescriptions for the infall and the SF rate.  In
  Table \ref{models} we summarize all the models we consider as a
  function of the type of infall, the timescale of the thin disk
  phase, the presence of the threshold, and the type of radial
  inflow. In particular, in the first column we report the model
  names, in the second the infall type, in the third in the forth it
  is specified if there is inside-out formation ($\tau_{D}$ variable
  with the radius) and the threshold in the SF,
  respectively. Finally, in the fifth it is indicated the type of the
  implemented radial flow (only for the ``no-static'' models).

 First of all, we present the results obtained with the one-infall
 model with an inside-out formation. In Fig. \ref{tut} we show 3
 different cases without radial flow: a model without threshold in the
 SF (S1I), models S1IT4 and S1T7 with a gas threshold for SF of 4 and
 7 $M_{\odot}$ $ pc^{-2}$, respectively,  compared with the data
   from HII and PNe. We conclude that even if we consider a model
 with an inside-out formation the obtained gradient with a one-infall
 model without threshold is too flat and the observational data are
 not reproduced especially in the outer part of the Galaxy disk. We
 see instead that the slope for galactocentric distances between 4 and
 14 kpc can be reproduced if we consider the model with a threshold in
 the SF. However, since the surface mass gas density in the outer
 parts of the disk is too small, there is no SF in this region and
 then no metals production and no chemical evolution.

\begin{figure}
\includegraphics[width=0.45\textwidth]{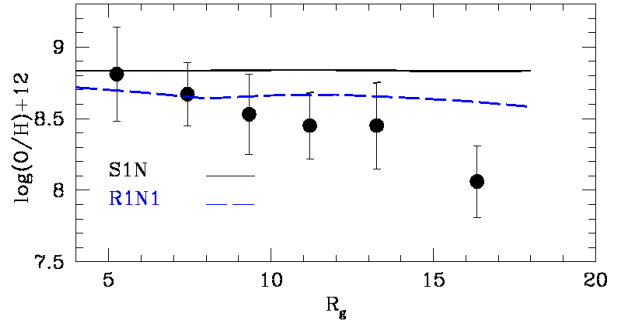}
\caption{Radial abundance gradient for oxygen. The black solid
  line refers to the one-infall model without threshold and with a
  constant $\tau_d=4$ Gyr (S1N). The blue long dashed line represents the
  model with a radial inflow of 1 km/s (R1N1). The data are the same of
  Fig. \ref{tut}. }
\label{t}
\end{figure}

 Therefore, a possible solution to reproduce the data can be to
  consider a radial inflow of gas.

  Portinari \& Chiosi (2000) computed the effects of radial flows on their
 one-infall model without an inside-out formation.  They adopted an
 unique timescale for disk formation irrespective  of the galactocentric
 distance.  Surprisingly, they found also in the case without radial flows an
 abundance gradient with a significant slope along the disk for the
 oxygen.

 In the Fig. \ref{t} we report the model  S1N using
 the same prescriptions of the SI1 model
 but without the inside-out formation, namely keeping
   $\tau_{D}$ constant along the Galaxy disk and equal to 4 Gyr. We
   find a flat gradient, and in the outer part of the disk the values
   of log(O/H)+12 are even larger than the value in the inner part of the
   Galaxy. Portinari \& Chiosi (2000) labeled S15a a model
   with the follow main prescriptions: i) one-infall, ii) no
   threshold, iii) no inside-out. With these prescriptions they found
   a significant gradient along the Galactic disk of $\sim -0.03$ dex
   kpc$^{-1}$ in contrast with our model result reported in
   Fig. {\ref{t} with the black solid line.
 
In Fig. {\ref{t} we also show the model R1N1 results where it was considered a constant speed of 1 km s$^{-1}$  for the radial inflow on the one-infall model no threshold and no inside-out formation.

\begin{table*}[htp]

\caption{The list of the models described in this work. The first letter in the model names is S if that model is a ``static'' one, otherwise  R if we show a model with radial flow of gas. The follow number (1 or 2) is related to the number of infall, the letter I is for the case for inside-out model and N for a constant formation time scale along the disk. When we consider a threshold in the star formation we label that with a T. The last letter in the names of models with radial flow is related to the speed of the radial flow. }
\scriptsize

\label{models}
\begin{center}
\begin{tabular}{c|ccccc}
  \hline
\noalign{\smallskip}

\\
 Models &Infall type& $\tau_d$& Threshold&Radial inflow  \\

  \\
\noalign{\smallskip}

\hline
\noalign{\smallskip}

S1I& 1 infall &  1.033 R (kpc) -1.27 Gyr &no&/\\
\noalign{\smallskip}
\hline
\noalign{\smallskip}
S1N & 1 infall  & 4 Gyr &no&/\\
\noalign{\smallskip}
\hline
\noalign{\smallskip}

S1IT4  & 1 infall &  1.033 R (kpc) -1.27 Gyr  & 4 M$_{\odot}$pc$^{-2}$&/\\
\noalign{\smallskip}
 \hline
\noalign{\smallskip}

S1IT7 & 1 infall  &  1.033 R (kpc) -1.27 Gyr   &  7 M$_{\odot}$pc$^{-2}$     &/\\

\noalign{\smallskip}

 \hline
\noalign{\smallskip}
 S2IT& 2 infall  &  1.033 R (kpc) -1.27 Gyr   & 7 M$_{\odot}$pc$^{-2}$ (thin disk phase) &/\\   
&      &   &4 M$_{\odot}$pc$^{-2}$ (halo thick disk phase)      &      \\

\noalign{\smallskip}

 \hline
\noalign{\smallskip}
 R1I1& 1 infall  &  1.033 R (kpc) -1.27 Gyr   &  no &- 1 km s$^{-1}$\\

\noalign{\smallskip}

 \hline
\noalign{\smallskip}
 R1N1 & 1 infall  & 4 Gyr &no&- 1 km s$^{-1}$\\
\noalign{\smallskip}

 \hline
\noalign{\smallskip}

 R1IL & 1 infall  &  1.033 R (kpc) -1.27 Gyr &no& linear inflow pattern\\
\noalign{\smallskip}

 \hline
\noalign{\smallskip}

R1IT01  & 1 infall &  1.033 R (kpc) -1.27 Gyr  & 4 M$_{\odot}$pc$^{-2}$&- 0.1 km s$^{-1}$\\
\noalign{\smallskip}

 \hline
\noalign{\smallskip}
R1IT05  & 1 infall &  1.033 R (kpc) -1.27 Gyr  & 4 M$_{\odot}$pc$^{-2}$&- 0.5 km s$^{-1}$\\
\noalign{\smallskip}

\noalign{\smallskip}

 \hline
\noalign{\smallskip}
 R2ITV& 2 infall  &  1.033 R (kpc) -1.27 Gyr   & 7 M$_{\odot}$pc$^{-2}$ (thin disk phase) &variable inflow pattern\\   
&      &   &4 M$_{\odot}$pc$^{-2}$ (halo thick disk phase)      &      \\

\noalign{\smallskip}

 \hline
\noalign{\smallskip}

\end{tabular}
\end{center}
\end{table*}

\begin{figure}
\includegraphics[width=0.45\textwidth]{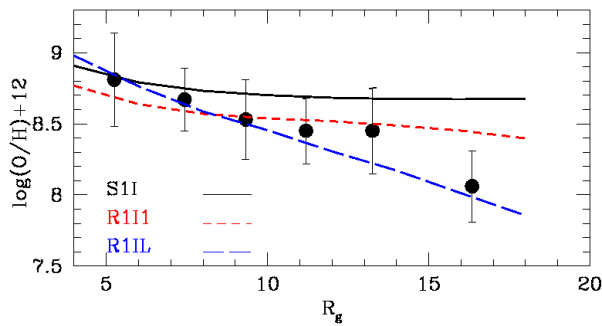}
\caption{Radial abundance gradient for oxygen. The black solid
  line refers to the one infall model without threshold and with the
  inside-out formation (S1I). The red short dashed line represents the model
  with a radial inflow of 1 km s$^{-1}$ (R1I1) and the blue long dashed line the
  best fit model using a variable velocity for the radial inflow (R1IL).  The
  data are the same of Fig. \ref{tut}}
\label{bestfit}
\end{figure} 

We see that the main effect of a migration of gas from the outer part
of the Galaxy toward the inner part without threshold and inside-out
is to produce a weak abundance gradient ($\sim$ -0.014 dex kpc$^{-1}$)
at variance with observations. In our model with a constant radial
velocity of the flow, the abundance of oxygen at each galactocentric
distance is lower than the values found for the S1N model; this is due
to the fact that the metals tend to be stored in the very central
parts (R $<$ 5 kpc) of the Galaxy and that we used a constant speed of
the radial flow. In fact, Sch\"onrich \& Binney (2009) considered both
the stellar and gas flows in a chemical evolution model considering a
model without inside-out formation. They were able to fit the observed
data including a variable speed of the gas flow by means of two free parameters, a situation where each ring may have its own velocity.
\begin{figure}
\includegraphics[width=0.45\textwidth]{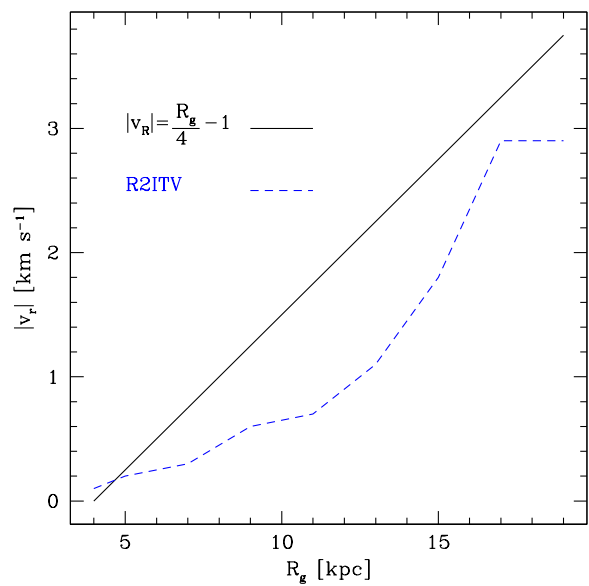}
\caption{ With the solid line we show the requested pattern of the
  velocity to reproduce the observed gradient as a function of the
  galactocentric distance for the R1IL model, with the dashed line for
  the R2ITV. 
In this plot we show the modulus of the radial inflow
  velocity as a function of the galactocentric distance.  }
\label{vel_1inf}
\end{figure}

\begin{figure}
\includegraphics[width=0.45\textwidth]{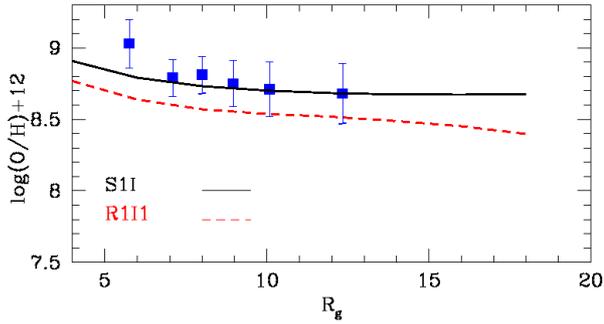}
\caption{Radial abundance gradient for  oxygen. The black solid
  line refers to the one infall model without threshold and with the
  inside-out formation (S1I). The red short dashed line represents the model
  with a radial inflow of 1 km s$^{-1}$ (R1I1). The data are mean values and relative errors from Cepheids by Andrievsky et al. (2002a,b).}
\label{cef1}
\end{figure}

We remind that also the S1I model with inside-out formation using the
law described in eq.  (\ref{t1}) and no threshold is not able to
reproduce the set of data for the outer part of the disk. Then we
tried to find a best fit to the observed data varying the velocity of
the gas flow. In Fig. \ref{bestfit} we report the R1I1 model results
with a constant radial flow along the Galactic disk fixed at -1 km
s$^{-1}$ and model  with an inflow velocity variable in space} which results
to be R1IL. We see that the model with the constant radial flow has an
steeper gradient compared to S1I but there are still problems in
reproducing the outer region of the Galaxy.  To reproduce the data we
need a variable velocity for the inflow of gas.  With the model
  RI1TL we are able to fit quite well the data of HII regions and PNe
  assuming a linearly increasing flow velocity towards the outer regions as shown in Fig. (\ref{vel_1inf}).  We recall that the case of a  linearly increasing flow velocity was also studied in the work of Goetz \& K\"oppen (1992).

\begin{figure}
\includegraphics[width=0.45\textwidth]{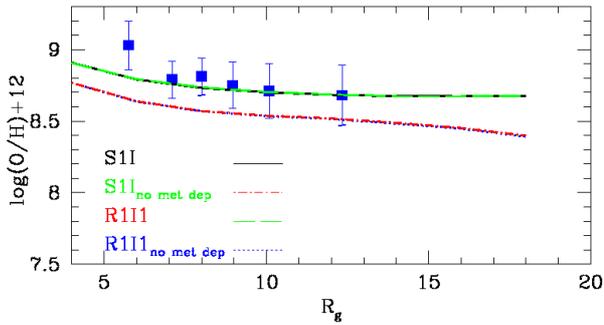}
\caption{Radial abundance gradient for oxygen. The black solid
  line refers to the one infall model without threshold and with the
  inside-out formation (S1I) with metallicity dependent yields, the green
  long dashed line is the case with solar yields . The red short
  dashed dotted line represents the model with a radial inflow of 1 km
  s$^{-1}$ (R1I1) with metallicity dependent yields and the blue long dashed
  line with solar yields. The data are mean values and relative errors
  from Cepheids by Andrievsky et al. (2002a,b).}
\label{metdep}
\end{figure}

\begin{figure}
\includegraphics[width=0.45\textwidth]{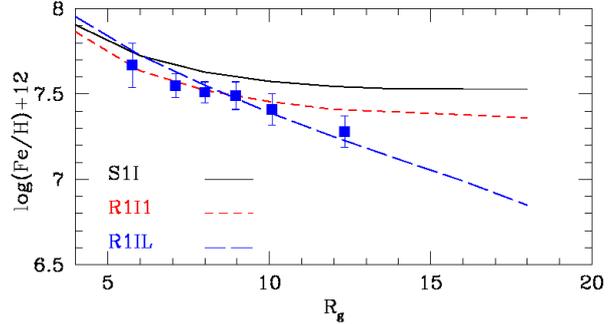}
\caption{Radial abundance gradient for  iron. The black solid line
  refers to the one infall model without threshold and with the
  inside-out formation (S1I). The red short dashed line represents the
  model with a radial inflow of 1 km s$^{-1}$ (R1I1,) and the blue
  long dashed line the best fit model using a variable velocity for
  the radial inflow (R1IL).  The mean values and relative errors from
  Cepheids by Andrievsky et al. (2002a,b) are reported with filled
  blue squares.}
\label{fe}
\end{figure}

\begin{figure}
\includegraphics[width=0.45\textwidth]{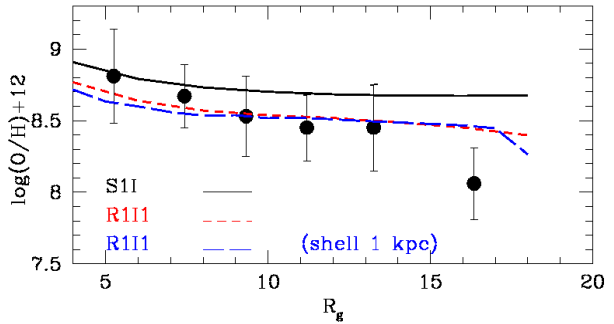}
\caption{Radial abundance gradient for oxygen. The black solid
  line refers to the one infall model without threshold and with the
  inside-out formation (S1I). The red short dashed line represents the
  model with a radial inflow of 1 km s$^{-1}$ (R1I1) using shells 2
  kpc wide and the blue long dashed line label the same physical
  parameters but running a model with shells 1 kpc wide. The data are
  mean values and relative errors from Cepheids by Andriesky et
  al. (2005a,b).}
\label{1kpc}
\end{figure}

\begin{figure}
\includegraphics[width=0.45\textwidth]{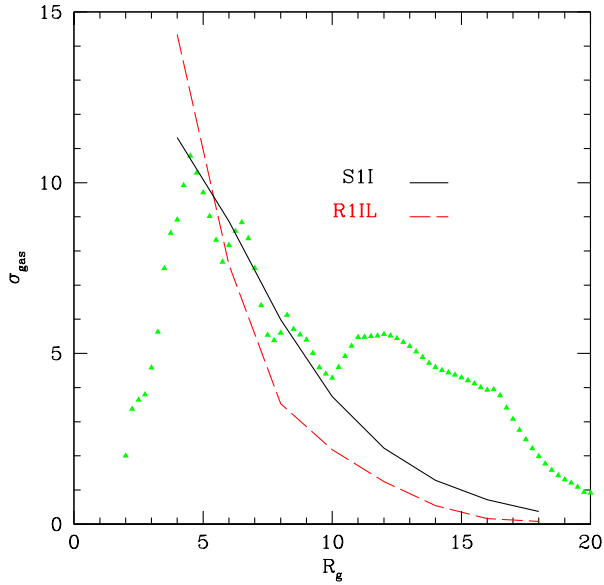}
\caption{ The surface gas density profile as a function of the radial
  distance for the reference one-infall model S1I drawn with the black
  solid line and for the best fit model R1IL. The green filled
  triangles represents the data of Dame et al (1993).}
\label{gas}
\end{figure}

 In Fig.  \ref{vel_1inf} we show the pattern of the  linear
   velocity we adopted to reproduce the observed gradient as a function
   of the galactocentric distance for the one-infall model. The range of
   velocities in modulus span the range 0-4 km s$^{-1}$ in
   accordance with the results of Sch\"onrich \& Binney (2009).

\begin{figure}
\includegraphics[width=0.45\textwidth]{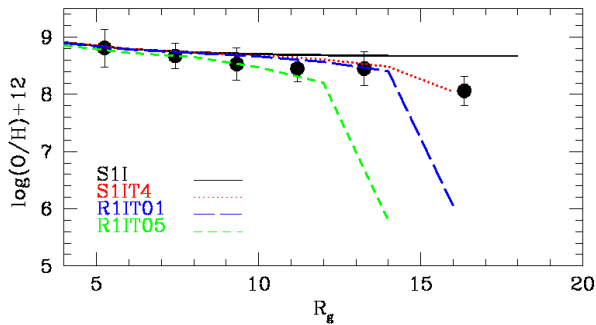}
\caption{Radial abundance gradient for oxygen. The black solid
  line refers to the one-infall model without threshold and with the
  inside-out formation (S1I). The red short dashed line represents the
  model with a threshold of 4 $M_{\odot} pc^{-2}$ (S1IT4), the blue
  long dashed line the model with a threshold of 4 $M_{\odot} pc^{-2}$
  combined to a radial inflow of 0.1 km s$^{-1}$ (R1T01), the green
  dotted line the model with a threshold of 4 $M_{\odot} pc^{-2}$
  combined inflow of 0.5 km/s (R1T05). The data are the same of
  Fig. \ref{tut}.}
\label{mix}
\end{figure} 

\begin{figure}
\includegraphics[width=0.45\textwidth]{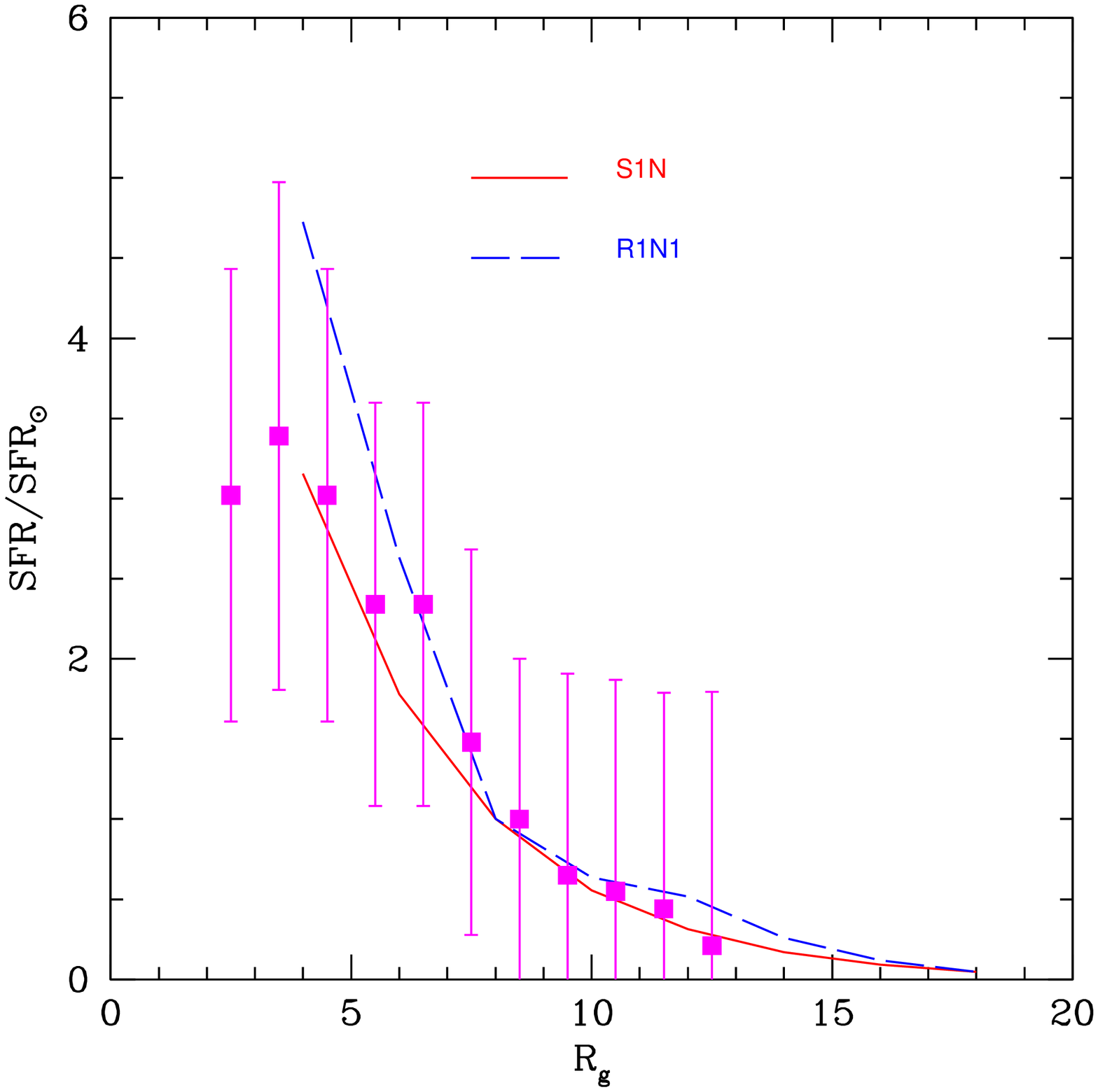}
\caption{The SF rate normalized to the solar values as a function of the
  galactocentric distance for the one-infall model with no inside-out
  formation. The data are taken by Rana et al. (1991) and are reported
  with the magenta points and relative error bars, with the red solid line we report the
  reference static model S1N whereas the dashed blue line is the model
  with a radial inflow of -1 km s$^{-1}$ (R1N1). }
\label{sfr1}
\end{figure} 

\begin{figure}
\includegraphics[width=0.45\textwidth]{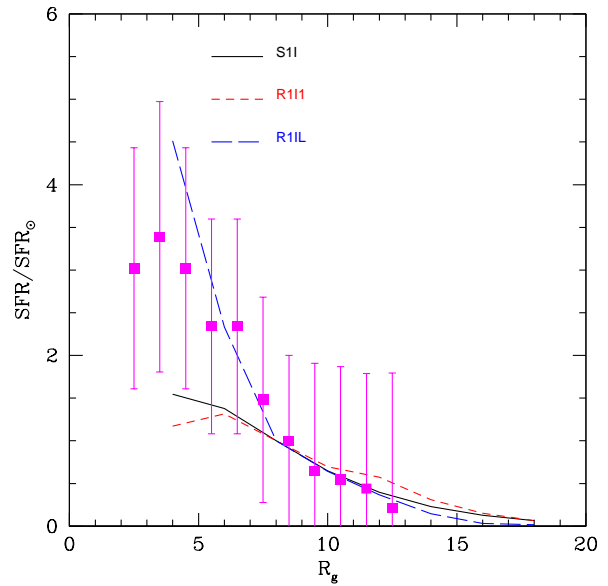}
\caption{The SF rate normalized to the solar values as a function of the
  galactocentric distance for the one-infall model with inside-out
  formation. The data are taken by Rana et al. (1991) and are reported
  with the magenta points and relative error bars, with the black solid line we report the
  reference static model S1I whereas the  red short dashed line is the
  model with a radial inflow of -1 km s$^{-1}$ (R1I1) and the blue long dashed line
  is the best fit radial inflow model (R1IL).}
\label{sfr2}
\end{figure} 
  
 In Fig. \ref{cef1} we compare the models S1I and R1I1 with the data
  from Cepheids. We note that, since the observed gradient for these
  objects is almost flat, both models can easily reproduce it.

 We recall that in our calculations we consider the nucleosynthesis
 prescription used in Fran\c cois et al. (2004), therefore for oxygen
 we use  metallicity dependent yields. To ascertain how the choice of metal
 dependent yields affects radial flows, in Fig. \ref{metdep} we run
 the same models of Fig. \ref{cef1} but in the case of solar yields
 for oxygen. We note that the reference models as well as the ones
 with radial flows, give exactly the same results for metal dependent
 and solar yields concerning the abundance gradients. This result
   is expected since the present time abundances in the ISM are the
   result of the global production of a given element; metal dependent
   yields vary with the metallicity but the global production of an
   element is the same as for solar metallicity yields since in both
   cases the solar abundances(i.e. the ISM abundances 4.5 Gyr ago) are
   reproduced.

 In Fig. \ref{fe} we show the results of models S1I, R1I1, and R1IL
 for iron.  We note that iron shows a steeper gradient when
 compared with oxygen one, because of the different time-scales
 involved in the  production of these two elements.

In our calculations we assume that: i) there is no inflow from the
outer parts of the disk ($R>20$ kpc), ii) our shells are 2 kpc wide.

The point ii) is consistent with the fact that on smaller scales other
processes must be important. In fact, in Spitoni et al. (2008) we
proved that galactic fountains can affect abundance gradients only on
scales smaller than 1 kpc. Here we test how the choice of a finer
resolution affects our results. In Fig. \ref{1kpc} the model R1I1 is
considered both for shells 1 and 2 kpc wide. We note that differences
arise only in the outer part of the disk, because of the adopted
method for the implementation of radial inflows. It is worth noting
that choosing a smaller shell does not affect our results expect for
distances R$>$ 17 kpc and not in a substantial way as shown in
Fig. \ref{1kpc}.

We pass now to analyze how the profile of the gas along the Galactic
disk is modified with the variable velocity pattern used for the best
fit of the abundance gradient. In the Fig. \ref{gas} we compare the
results of model S1I and R1IL for what concerns the gas surface
density profile as a function of the radial distance.

We note that both models underestimates the gas for distances $>$ 10
kpc and do not reproduce the peak and the decrease of gas for R $<5$
kpc. This is probably due to the bar which is not considered in our
model.

We
want also to stress that a model without a threshold in the SF
has problems to reproduce the values in outer part of the disk.
\begin{figure}
\includegraphics[width=0.45\textwidth]{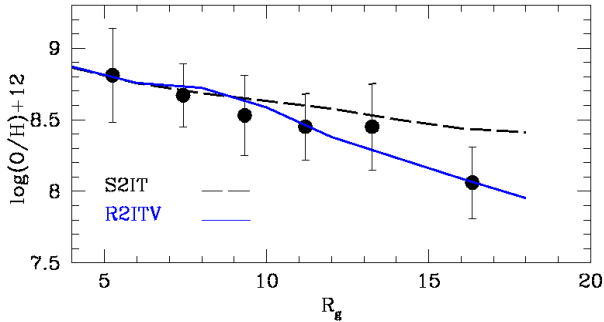}
\caption{Radial abundance gradient for oxygen. The black long
  dashed line refers to the two-infall model S2IT. The
  blue solid line represents the the best 2 infall fit model using a
  variable velocity for the radial inflow (R2ITV).  The data are the same of
  Fig. \ref{tut}}
\label{2infall}
\end{figure}
We also test a model in which we combine the effect of a threshold in
the SF plus a radial inflow. In Fig. \ref{mix} is shown
the ``static ``  model S1IT4 with the ``mixed models''
considering both the threshold and the radial inflow. We 
fixed  a threshold value at 4 $M_{\odot}$ pc$^{-2}$ considering 2
constant speed  radial inflows: the model RI1T01 with a velocity of 0.1 km
s$^{-1}$ and the model RI1T05 with a velocity 0.5 km s$^{-1}$,
respectively. A velocity of 0.5 km/s is
high enough to put system under threshold already at the radial
distance of 16 kpc, whereas the model without radial flow but with a
threshold of 4 $M_{\odot}$ pc$^{-2}$ has no chemical evolution for R $>$ 16 kpc.


 We want also to test how the radial inflows of gas affect the SF rate
 along the disk of the Galaxy. In Fig. \ref{sfr1} we report the
 results for the S1N and the R1N1 models, whereas in Fig.  \ref{sfr2}
 the results with inside-out formation: S1I, R1I1 and R1IL models. In
 these Figs. we show the SF rate normalized to the solar value as a
 function of the galactocentric distance. Because of the
   uncertainties in this data set, as evident from the big error bars,
   we cannot draw firm conclusions. Having said that, we can however
   draw some considerations. In Fig. \ref{sfr1} both models well fit
 the data for the SF rate, whereas in Fig.  \ref{sfr2} models SI1 and
 R1I1 underestimate the SFR/SFR$_{\odot}$ for R $<$ 8 kpc. On the
   other hand, the model R1IL fits perfectly the SFR/SFR$_{\odot}$
   ratio in the range 6 - 18 kpc. For galactocentric distances smaller
   than 6 kpc the R1IL model overestimates the mean SFR/SFR$_{\odot}$,
   but our results are inside the 1 sigma error bar, because of the
   uncertainty in the data.  We do not show our model results for
   galactocentric distances smaller than 4 kpc because in this region
   the effects of the central bar might be important and our model is
   built only for the disk and not for the bulge area.

In this part we study the effects of a radial inflow on the two-infall
model described in Sect. 2. In Fig. \ref{2infall} we see that our
two-infall reference model S2IT is able to give rise to a steep
gradient for oxygen but not enough to reproduce the data set. We also
show the model R2ITV with a varying velocity pattern for the radial
inflow velocity.  In the case of the two-infall model we are not
  able anymore to fit the abundance gradient for oxygen using a simple
  constant relation between radial infall velocity and galactocentric
  distance, then we consider a pattern where the velocity starting
  from 17 kpc decreases monotonically toward the center part of the
  Galaxy (see Fig. \ref{vel_1inf}).

From Fig. \ref {vel_1inf} we note that
the range of velocities required to fit the data is quite similar to
the one used for the one-infall model.

\begin{figure}
\includegraphics[width=0.45\textwidth]{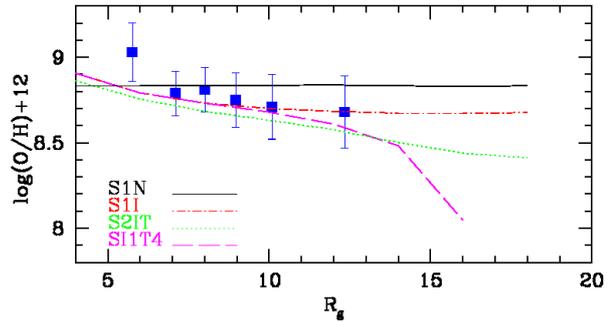}
\caption{Radial abundance gradient for oxygen for the models S1N
  (black solid line), S1I (red dotted dashed line), and S2IT (green
  dotted line), and SI1T4 (magenta long dashed line). The data
  collection from Cepheids are reported.}
\label{ce}
\end{figure} 

 In Fig. \ref{ce} we report the data from Cepheids compared with
 standard models without radial flows (S1N, S1I, SI1T4, S2IT). We see
 that all the models with inside-out formation show a rather shallow
 slope for the oxygen abundance gradient compatible with the Cepheid
 data.  In Fig. \ref{ce2} we compare standard models without radial
 flows with all data sets considered in this paper. In conclusion, if we use only the data of the
 Cepheids, the abundance gradient is not so steep and it can be
 reproduced without any radial inflow.
\begin{figure}
\includegraphics[width=0.45\textwidth]{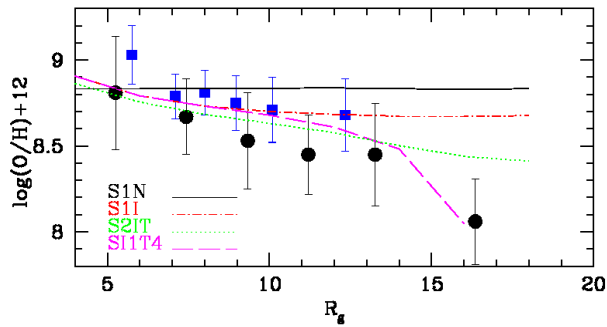}
\caption{Radial abundance gradient for oxygen for the models S1N
  (black solid line), S1I (red dotted dashed line), and S2IT (green
  dotted line), and SI1T4 (magenta long dashed line). The data
  collection from Cepheids (blue squares ) and from HII regions (black circles) are reported.}
\label{ce2}
\end{figure}

\section{Conclusions}

In this paper we have studied the effects of radial inflows of gas on
a detailed chemical evolution model. We also reported some results of
the metals gradients obtained with ``static'' models considering
different prescriptions for the infall law and star formation. Our
main conclusions can be summarized as follows:

\begin{itemize}

\item{If we consider the one-infall model with an inside-out formation ($\tau_{D}$ varying with the radius)
  the obtained gradient without threshold is too flat and the
  observational data are not reproduced especially in the outer part
  of the Galaxy disk. The slope for intermediate galactocentric
  distances can be reproduced if we consider the model with a
  threshold. However, since the surface gas mass density in the outer
  part of the disk is too small, there is no star formation in this
  region and then no metals production and no chemical evolution.  }

\item{Taking into account a constant $\tau_{D}$ along the Galaxy disk
  equal to 4 Gyr, we find a flat gradient in
  contrast with the work of  Portinari \& Chiosi (2000) where
  a gradient along the Galactic disk of $\sim -0.03$ dex kpc$^{-1}$ was found.}

\item{A constant radial inflow with speed of 1 km s$^{-1}$ (not expected theoretically) applied to
  the one-infall model with no threshold and no inside-out produces a
  weak abundance gradient ($\sim$ -0.014 dex kpc$^{-1}$) and the
  metals tend to be stored in the very central parts of the Galaxy.}

\item{The required pattern of the the radial flow velocity varies
  linearly with the galactocentric distance and spans a range between
  0 and 4 km/sec. This conclusion holds for the one-infall model
  without threshold and with inside-out formation. }

\item{Our ``static'' two-infall model with a threshold in the star
  formation and the inside-out formation is able to give rise to a
  steep gradient but not enough to reproduce the data set. The
  required pattern of the radial inflow velocity (in modulus) to
  reproduce the observed gradient is quite similar to the one of the
  one-infall model. }

\item{If we use the data of the Cepheids the observed abundance
  gradient is not so steep and can be reproduced with our ``static''
  reference models without any radial inflow. }

\item {As shown first in G\"otz \& K\"oppen (1992),  then in Portinari \& Chiosi (1999) and Colavitti et al. (2008),
  a variable efficiency in
  the star formation rate if coupled with an inside-out formation can
  also reproduce the abundance gradient. Since this variable
  efficiency in the Milky Way is observationally motivated
  (Marcon-Uchida et al. 2010),  it should not be disregarded as source of the abundance gradients. Moreover, as we have shown, to
  reproduce the abundance gradients by means of radial flows we need a specific
  variable speed of the flow, and at this moment there are not
  observational constraints for the speed of this flow of gas along
  the disk of the Galaxy.  }

\item{Finally, we conclude that radial gas flows can be in principle
  important to reproduce the gradients along the disk, although an
  inside-out formation coupled with variable efficiency of star
  formation and threshold in gas density can also well reproduce the
  data without radial flows. Probably, all these processes are present
  in the Galactic disk: in order to decide which of these processes
  would prevail in the formation of the disk, we need more detailed
  data both on the abundance, gas and star formation rate gradients,
  as well as data on high redshift disks.  }

\end{itemize}

\begin{acknowledgements}
We thank the referee for the enlightening suggestions.  We also thank
G. Cescutti, and S. Recchi for many useful discussions. We aknowledge financial support from PRIN 2007 MUR Prot. No. 2007JJC53X-001.
\end{acknowledgements}

\end{document}